\documentstyle[procsla,epsfig,12pt]{article}
\begin{document}


\begin{titlepage} 

\vspace*{1.5truecm}

\begin{flushright}
\begin{tabular}{l}
NORDITA--98--1 P\\
hep-ph/9801222
\end{tabular}
\end{flushright}
 
\vspace{2cm}
 
\begin{center}
{\Large\bf Light-Cone Sum Rules}

\end{center}

\vspace{1.2cm}
 
\begin{center}
V.M.\ Braun\footnote{On leave of absence from
    St.\ Petersburg Nuclear Physics Institute, 188350 Gatchina, Russia.}\\
{\sl NORDITA, Blegdamsvej 17, DK--2100 Copenhagen \O, Denmark}
\end{center}
 
\vspace{1.5cm}

\begin{center}
{\bf Abstract}\\[0.3cm]
\parbox{12.5cm}{
    I give a short introduction to the method of light-cone sum rules,
their theoretical background and their main modifications.
The discussion is concentrated but not restricted to the 
applications to heavy quark decays.
               }
\end{center}
 
\vspace{2cm}
 
\begin{center}
{\sl Plenary talk given at the\\
IVth International Workshop on Progress in Heavy Quark Physics\\
Rostock, Germany, 20--22 September 1997\\
To appear in the Proceedings}
\end{center}

\end{titlepage}
 
\thispagestyle{empty}
 
\setcounter{page}{1}
 

{\Large\bf Light-Cone Sum Rules}\\[2mm]
{\it V.M.~Braun}\\[2mm]
{\small NORDITA, Blegdamsvej 17, DK--2100 Copenhagen \O, Denmark}\\[2mm]

\medskip
\noindent{\large\bf Introduction}
\medskip

Already a long time ago it was realized that large momentum transfer to an 
extended object (hadron) requires a specific configuration of its constituents.
One possibility is to pick up a configuration in which almost all
momentum is carried by one parton. The large momentum 
can be transferred to this fast parton which eventually  recombines 
with the soft cloud. The second possibility is to pick up the Fock state 
with a minimum number of constituents (quark and antiquark for a meson) 
at small transverse separations, and exchange a hard gluon.
In the first case the contributing transverse
distances are not restricted which makes this mechanism difficult for
theory; in the second case a factorizaton formula can be 
derived\,\cite{BLreport} and the relevant nonperturbative information can be  
parametrized by hadron distribution amplitudes given by vacuum-to-meson
matrix elements of light-cone operators.

Which mechanism actually dominates the cross section --- this is a nontrivial 
question which has to be studied case by case. For pion 
electromagnetic form factor it has been proven\,\cite{exclusive} that 
in the theoretical limit $Q^2\to\infty$ the 
``soft'' (or ``end-point'') contribution is suppressed 
compared to the ``hard'' contribution by one power of $1/Q^2$. 
For heavy-to-light B decay form factors at large recoil (e.g. $B\to \pi e\bar\nu$)
both soft and hard contributions are of the same order in the 
$1/m_b$ expansion. For practical values of $Q^2$ and $m_b$ the soft 
contribution is always numerically important and often dominates. 
Taking it into account is difficult and presents a notorious problem 
in the theory of hard exclusive processes, which is not solved until now. 

An important theoretical progress  which has allowed for  quantitative 
estimates of soft contributions was made with the arrival of QCD sum 
rules\,\cite{SVZ}. Within this approach,  
matrix elements of a certain operator 
$J$ sandwiched between two hadron states $h_1$ and $h_2$ can be evaluated
by studying  correlation functions of the type
\begin{equation}
 \!\int \!\! dx\, dy\, e^{-ip_1x+i p_2 y} 
\langle 0| T\{H_2(y) J(0) H_1(x)\}|0\rangle \sim
\langle 0|H_2|h_2\rangle \frac{1}{m_2^2-p_2^2} 
\langle h_2|J| h_1\rangle \frac{1}{m_1^2-p_1^2} 
\langle h_1 |H_1 |0\rangle
\label{SVZ}
\end{equation}
where $H_1$ and $H_2$ are suitable interpolation currents. The idea is
to make a matching 
between the short-distance expansion in Euclidian space and the expansion in
hadron states in the two variables $p_1^2$ and $p_2^2$ with fixed value of
$q^2=(p_2-p_1)^2$.
The detailed procedures have been worked out in Ref.\,\cite{3ptSR}
and involve double dispersion relations, double Borel 
transformation to suppress contributions of higher states, 
and using vacuum condensates\,\cite{SVZ} to take into account 
nonperturbative effects. 

The case $q^2=(p_2-p_1)^2 =0$ is special and requires a certain modification
of the operator product expansion (OPE) to include so-called bilocal
power corrections corresponding to contributions of large distances
in the ``t-channel'' (the region of large $x$ and $y$ in (\ref{SVZ}) such that
$|x-y| \sim 1/|p_{1,2}| \to 0$)\,\cite{bilocal}. 
The structure of such modified OPE is well understood if one keeps
$p_1=p_2 \equiv p$ identically. Because of this restriction, the procedure 
is somewhat different compared to the form factor case: One  uses
an ordinary dispersion
relation in the single remaining variable and finds contribution of 
interest as the one which
multiplies a double-pole term $\sim 1/(m^2-p^2)^2$.  

This extension of the original SVZ sum rules to three-point functions
has proved
 to be quite successful and has a lot of applications, for example to
pion and nucleon form factors at intermediate momentum transfers, to
semileptonic form factors of D decays, to baryon magnetic moments and 
axial constants, to $g_{\pi NN}$ and $g_{\pi B B^*}$ couplings, 
and to many other physical observables.  

The increase in sophistication has its price, however. The three-point sum 
rules have specific problems which severely restrict their potential accuracy
and region of applicability. These subtleties 
are well known to experts, but very often escape due attention of the 
majority of sum rules ``users'' and the physics community in general.

\medskip
\noindent{\large\bf Problems of Three-Point Sum Rules}
\medskip

The first major problem is that 
\begin{itemize}
  \item OPE (short-distance expansion in condensates) upsets  power 
     counting in the large momentum/mass.
\end{itemize}
This problem is known since already the first sum rules for 
the pion electromagnetic form factor\,\cite{3ptSR} which
have the following (schematic) structure:
\begin{equation}
 F_\pi(Q^2) \sim \#\cdot \frac{1}{Q^2} 
 +\#\cdot \frac{\langle g^2G^2\rangle}{M^4}
 +\#\cdot Q^2 \frac{\langle \bar q q\rangle^2}{M^8}+\ldots
\end{equation}
Here $M^2$ is the Borel parameter which is of order 1 GeV$^2$.
The first contribution is due to  perturbation theory and it has 
the expected $1/Q^2$ behavior\footnote{The leading-order term is only
$\sim 1/Q^4$ 
and the $1/Q^2$ behaviour starts with the radiative correction.}.
 Contribution of the gluon condensate 
is independent of $Q^2$ and it is easy to convince oneself that
condensates of higher dimension are accompanied by 
increasing powers of $Q^2$. If plotted as a function of $Q^2$, 
the sum rule result for $F(Q^2)$ starts to {\em rise} at 
$Q^2\geq 3-5$ GeV$^2$. Such behavior is clearly unphysical and 
indicates that at high momentum transfers the OPE breaks down.
Requiring that contributions of higher dimension constitute a moderate 
fraction of the perturbative result, one obtains that 
the sum rule is only legitimate in a narrow interval of 
$0.5 \leq Q^2 \leq 1.5$ GeV$^2$.
 (The lower limit is due to the neglect of bilocal 
power corrections in this approximation.) 

The three-point sum rules for heavy-to-light decays have the similar problem
at large recoil. For example, the sum rule for the form factor $A_1$ in 
$B\to\rho e\bar\nu$ at the maximum recoil $q^2=0$ has the following 
structure:
\begin{equation}
A_1(q^2=0) \sim \# \cdot\frac{1}{m_b^{3/2}} +
                \#\cdot m_b^{1/2} \langle \bar q q\rangle + 
                 \#\cdot m_b^{3/2} \langle \bar qg\sigma G q\rangle + \ldots
\end{equation}
Here $q^2$ is the invariant mass of the lepton pair. For generic values 
of $q^2$ the relevant large parameter is the  
energy of the outgoing $\rho$ meson $(m^2_b-q^2)/(2m_b)$ in the B rest frame,
which plays the role of the hard momentum transfer $Q$ in the above example.
The similarity is clear. The rise of the form factor $A_1$ observed in
calculations using three-point sum rules is entirely due to this 
principal problem: expansion in slowly varying (vacuum) fields is
inadequate if a short-distance subprocess is involved.  
For decays of D mesons\,\cite{BBD} the recoil energy is not large -
comparable to the region of applicability of sum rules for the pion 
form factor, and the traditional approach works well. 
For B decays it does not work
apart from the specific case of $B\to\pi e\bar\nu$  
transition\,\cite{BBD91,Ball93} where (accidentally)  the quark condensate  
contribution is only  $\sim m_b^{-1/2}$ 
and the problem is numerically less important.
 
\medskip

The second general problem of three-point sum rules is 

\begin{itemize}
 \item Contamination of the sum rule by ``nondiagonal''
  transitions of the ground state to excited states.
\end{itemize}
This is a notorious problem in calculations of hadron matrix elements 
at zero momentum transfer. The contribution of interest corresponds in
this case to the double-pole term in the correlation function
(\ref{SVZ}) at $p^2=m_h^2$
while transitions from the ground state to excited states generically
produce single-pole terms which are not suppressed by the Borel 
transformation:
\begin{equation}
  \frac{1}{(m^2_h-p^2)^2}\cdot \langle h|J|h\rangle +
  \frac{1}{(m^2_h-p^2)}\frac{1}{(m^2_h-m^2_{h'})}
\cdot \langle h|J|h'\rangle +\ldots
\end{equation}
In order to get rid of ``parasitic'' single-pole contributions one is 
forced to introduce additional parameters or take the derivative 
of the sum rule in respect to the Borel parameter, resulting in a 
considerable loss of accuracy. In addition, it becomes not possible 
to take into account  mass difference of the initial and final 
hadrons since one can rewrite 
\begin{equation}
 \frac{1}{(m^2_1-p^2)}\frac{1}{(m^2_2-p^2)}
  =  \frac{1}{(m^2_1-m^2_2)}\left[
 \frac{1}{(m^2_2-p^2)}-\frac{1}{(m^2_1-p^2)}\right]
\end{equation} 
and there is no double-pole term at all. This does not allow for 
calculations of transition matrix elements of the type $\Sigma\to p\gamma$,
$\Delta\to N\gamma$ etc., where mass differences are large.

{}For form factors at sufficient values of $q^2$ it was proposed\,\cite{3ptSR}
to get rid of nondiagonal transitions by using the double dispersion 
relation and taking  Borel transform in both variables. In practical 
applications there are 
several caveats, however: First, the results depend on the shape 
of the duality region in plane of the two dispersion variables 
and this dependence can be significant. Second, there are formal problems with
double dispersion relations in the decay kinematics in presence of
Landau singularities\,\cite{BBD}. Third, it is becoming increasingly clear
that  suppresion of nondiagonal transitions by the double Borel transform
is more formal than real\footnote{For example, nondiagonal transitions upset
QCD sum rules for the b quark kinetic energy in B meson, 
see Ref.\,\cite{BL97} and references therein.}.
 
\medskip
 
Light-cone sum rules (LCSR)\,\cite{BBK,BF1,CZ90} were developed in 
late 80-th in an  
attempt to solve or at least moderate the problems of three-point
sum rules by making a partial resummation of the OPE to all orders
and reorganizing the expansion in terms of twist of relevant operators
rather than their dimension\footnote{The term ``light-cone sum rules''
first appears in Ref.\,\cite{BBD}.}. In physical terms, the 
difference is that the expansion at short distances is substituted by 
the expansion in the {\em transverse} distance between partons in the 
infinite momentum frame. In this way one incorporates certain 
additional information on QCD correlation functions related to
approximate conformal symmetry of the theory. Technically, the LCSR 
approach presents a marriage of QCD sum rules with the
theory of hard exclusive processes. As a bonus, 
SVZ  vacuum condensates are substituted by 
light-cone hadron distribution functions of increasing twist which have
a direct physical significance.

\begin{figure}
\vspace*{-2.4cm}
\centerline{\epsfxsize10.0cm\epsffile{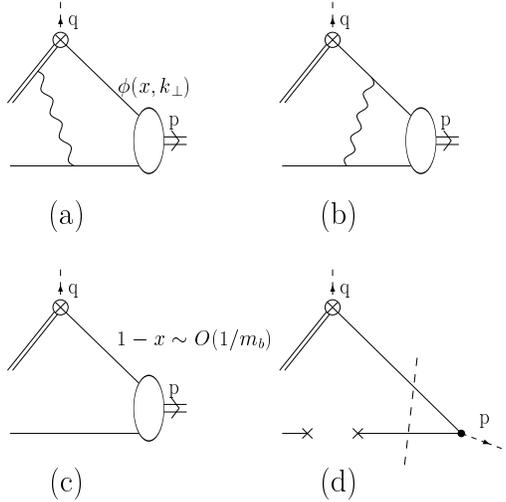}}
\vspace*{-5.5cm}
\caption[]{``Hard'' (a,b) and ``soft'' (c) contributions to the decay 
form factor. (d): modelling of the soft contribution in the QCD sum rule
approach.}
\end{figure}

\medskip
\noindent{\large\bf A Heuristic Discussion}
\medskip

Consider the semileptonic decay
$B\to\pi e\bar\nu$ at zero invariant mass of the lepton pair $q^2=0$, 
pictured schematically in Fig.~1. The $u$ quark in the final state 
has large energy of order $E_u\sim m_b/2$ in the B meson rest frame
and has to recombine with the soft spectator antiquark with $E_{\bar d}
\sim \Lambda_{\rm QCD}$ to form a pion. If no hard gluons are exchanged
as in Figs.~1a,b (we will discuss this ``hard''\,\cite{hard} 
contribution later), 
the form factor is proportional to the overlap integral of such an 
asymmetric configuration --- fast quark
and slow antiquark  --- with the pion state, see Fig.~1c.
 Schematically, we can write
\begin{equation}
   f_+^{B\to\pi}(q^2=0) \sim m_b^{1/2}\cdot
\!\!\!\!\!\!\!\!\int\limits_{1-O(1/m_b)}^1 
\hspace{-0.5cm} dx\,\phi_\pi(x,b)
\label{ffac}
\end{equation}
where $x=2E_u/m_B$ is the $u$ quark energy fraction and $b$ is the 
separation between the quark and the antiquark in the plane transverse
to the (large) pion momentum\footnote{It can be argued that contributions
of other than valence states (with additional gluons and/or $q\bar q$ pairs)
are suppressed by extra powers of $1/m_b$.}. The extra factor $m_b^{1/2}$
is due to the normalization of the B meson coupling to the $b\bar d$
pair; it is not very important for what follows. 

{}For sufficiently small $b$ one can derive the asymptotic behavior 
of the pion distribution amplitude $\phi_\pi(x,b)$ at large x
\begin{equation}
  \phi_\pi(x,b\ll 1/\Lambda_{QCD}) \stackrel{x\to 1}{=} - N(b)\cdot (1-x)
\label{largex}
\end{equation}   
where the $b$-dependent normalization factor $N(b)$ 
\begin{equation}
 N(b) = 6\left[1+ 
   6\, a_2(b_0) \left(\frac{\alpha_s(b)}{\alpha_s(b_0)}\right)^{50/81}
 + 15\, a_4(b_0) \left(\frac{\alpha_s(b)}{\alpha_s(b_0)}\right)^{364/405}
 +\ldots\right]
\end{equation}
can be calculated in terms of (nonperturbative) coefficients $a_n$ at 
a certain reference scale. Provided that this expansion is convergent  ---
which certainly is the case at very small transverse separations  ---
the behaviour $\phi(x)\sim (1-x)$ is maintained by the renormalization group
evolution.
Assuming (\ref{largex}), the contribution of small transverse 
separations to the form factor scales as\,\cite{CZ90,ABS} 
\begin{equation}
            f_+^{B\to\pi}(q^2=0) \sim m_b^{1/2}\cdot \frac{1}{m_b^2} = 
            \frac{1}{m_b^{3/2}}\,.
\label{3/2}
\end{equation} 
My discussion was purely heuristic. It can be made more rigorous with the
result that  
(\ref{3/2})  is indeed the correct behavior in the theoretical 
limit $m_b\to\infty$, by observing\,\cite{ASY94} that contributions 
of large transverse separations are suppressed by Sudakov effects.
Problem is, however, that the Sudakov suppression is very weak.
With the b quark mass of order 5 GeV it 
becomes effective at $b\sim 1$~fm only, which is 
of order or even larger than the B meson radius 
determined by nonperturbative effects. 
Taking them into account is mandatory for a quantitative analysis.

It is here that ideas of the 
 QCD sum rules enter the stage: I will try to make a {\em matching}
between the QCD calculation at small $b$ with the expansion in hadron 
states at large transverse separations.
If the behaviour in (\ref{largex}) is correct at relatively low scales 
of order 1 GeV where the matching is made to hadronic states
--- which is supported by the existing evidence, see below --- then the
power counting in the quark mass (\ref{3/2}) is correct for realistic values 
of the b quark mass at which the perturbative (Sudakov) dominance
of small impact parameters does not hold yet. 

It is instructive to explain in this
language why traditional three-point sum rules fail to describe 
B meson decays. The leading nonperturbative effect
is then given by the diagram in Fig.~1(d), where the light quark
is soft and interacts with the nonperturbative QCD vacuum, forming the
quark condensate. Since quarks in a condensate
have zero momentum, this diagram yields a
contribution to the distribution amplitude that
is naively proportional to $\delta(1-x)$. The corresponding contribution to
the decay form factor (\ref{ffac})  remains
unsuppressed for $m_b\to\infty$ and obviously 
violates the power counting discussed above. 
The contradiction must be resolved by including the contributions of
higher-order condensates to the sum rules and subtracting the 
contribution of excited states.
The suppression of the end-point region $x\to 1$, which is
expected in  QCD, can only hold as a {\em numerical}
cancellation between different contributions, which becomes the more
delicate (and requires more fine-tuning) the more $m_b$ increases.
{}For $m_b\approx 5\,$GeV a suppression of
the quark condensate contribution by a factor
$\sim 1\,$GeV$^2/m_b^2\sim 1/25$ is required. 
This explains why the three-point sum rules
become unreliable. 
 
\medskip
\noindent{\large\bf A Simple Light-Cone Sum Rule}
\medskip

After these preliminary remarks, I will now derive the simplest LCSR
for the $B\to\pi e\bar\nu$ form factor\,\cite{CZ90,BBD,BKR93,Bel95}.
To this end, consider the correlation function
\begin{equation}
\Pi_\mu(p_B^2,q^2) = i\int\!d^4z\,e^{-ip_Bz}
\langle \pi(p_\pi)|T\{\bar u(0)\gamma_\mu b(0) b(z)i\gamma_5 d(z)\}|0\rangle.
\label{B-Pi}
\end{equation}
At large negative $m_b^2-p_B^2$ and fixed (small and positive) $q^2$
this correlation function can systematically be calculated in QCD.
The leading contribution is expressed in terms of the pion distribution 
amplitude:
\begin{equation}
 \Pi_\mu(p_B^2,q^2) = (p_B-q)_\mu f_\pi m_b
  \int\limits_0^1 \!dx\, \frac{\phi_\pi(x,\mu)}{m^2_b-x p_B^2-(1-x)q^2}
  +\ldots
\label{lt}
\end{equation}
where $\mu^2\sim m^2_b-p_B^2$
and the corrections are suppressed either by powers of $\alpha_s(\mu)$
(radiative corrections) or by powers of $1/(m_b^2-p_B^2)$ (higher twist 
corrections)\footnote{From now on I will set the scale in the distribution
amplitude by the cutoff in the transverse momentum rather than position
space, which is more convenient in practical calculations.}. 
On the other hand, $\Pi_\mu(p_B^2,q^2)$ has a pole at $p_B^2=m_B^2$ 
corresponding to the B meson intermediate state:
\begin{equation}
\Pi_\mu^{\rm B meson}(p_B^2,q^2) =  \frac{f_B m_B^2}{m_b}\cdot
 \frac{1}{m_B^2-p_B^2}\cdot 
\left[(2p_B+q)_\mu \,f_+^{B\to\pi}(q^2)+q_\mu\, f_-^{B\to\pi}(q^2)\right] .
\label{pole}
\end{equation}
We can relate the two above representations, observing that $\Pi(p_B^2,q^2)$
is analytic in the cut $p_B^2$ plane, and assuming that the B meson
contribution is given by integral of the QCD spectral density
over the {\em interval of duality} $m_b^2 < s < s_0$:
\begin{equation}
  \Pi_\mu^{\rm B meson}(p_B^2,q^2) = (p_B-q)_\mu 
   \int\limits_{m_b^2}^{s_0}\!\frac{ds}{s-p_B^2}\, \rho(s,q^2).
\label{duality}
\end{equation} 
The explicit expression for $\rho(s,q^2)$ can easily be read off (\ref{lt}),
making a change of variables $x\to s = (m^2_b -q^2)/x+q^2$.
Equating Eqs.~(\ref{pole}) and (\ref{duality}) and making 
the  Borel transformation $(s-p_B^2)^{-1}\to \exp(-s/M^2)$,
$(m_B^2-p_B^2)^{-1}\to \exp(-m_B^2/M^2)$, we obtain (after some 
rewriting) the {\em light-cone sum rule}
\begin{eqnarray}
 \frac{f_B m_B^2}{f_\pi m_b}f_+^{B\to\pi}(q^2) e^{-(m_B^2-m_b^2)/M^2}&=&
\frac{1}{2}\int\limits_{x_0}^1\!\frac{dx}{x}\,\phi_\pi(x,\mu)
 \,e^{-\frac{(1-x)(m_b^2-q^2)}{xM^2}},
\nonumber\\
x_0 &\equiv& \frac{m_b^2-q^2}{s_0-q^2}.
\label{LCSR1}
\end{eqnarray}
Note that the restriction in the maximum invariant mass of the heavy-light
quark pair $s<s_0$ translates to the lower limit in  the momentum
fraction carried by the b quark $x>x_0$. In the heavy quark limit
$s_0\simeq (m_b+ 1$ GeV$)^2$ and $x_0\simeq 1- O(1/m_b)$ in agreement 
with the heuristic discussion above.  

Compared to the traditional three-point QCD sum rules, note:
(i) a single variable dispersion relation; (ii) no condensates; 
(iii) resummation of contributions of operators of leading twist.
The last statement follows from the definition of the pion distribution
amplitude: Moments of $\phi_\pi$  equal vacuum-to-pion matrix elements 
of twist two operators
\begin{equation}
\langle \pi(p)|\bar u\gamma_\nu \gamma_5 
  \stackrel{\leftrightarrow}{D}_{\mu_1}\ldots 
  \stackrel{\leftrightarrow}{D}_{\mu_{n-1}} d|0\rangle =
 -i f_\pi p_\nu p_{\mu_1}\ldots p_{\mu_{n-1}}
\int\limits_0^1\! dx\,(2x-1)^n\phi_\pi(x,\mu)
\end{equation}
Naively, each such matrix element presents an independent nonperturbative
parameter $M_n \equiv \int_0^1\! dx\,(2x-1)^n\phi_\pi(x)$. It is easy to 
check that expansion of the r.h.s. of the sum rule (\ref{LCSR1})
in moments $M_n$ would correspond to expansion of the distribution function
in derivatives of $\delta(1-x)$, which is the origin  
of problems with the traditional sum rules.
The crucial idea of the LCSR approach  
is that the expansion in moments, alias in operators of increasing dimension,
 is replaced by the expansion in conformal partial waves, each of which 
takes into account a subset of operators to all dimensions. 

The trick is analogous to the partial wave expansion
of the wave function in usual quantum mechanics.
The rotational symmetry of the potential allows one (in quantum mechanics)
to separate angular and radial degrees of freedom.
The dependence on the angular 
coordinates is included in spherical harmonics which form an irreducible
representation of the $O(3)$ group, and the dependence on the single 
remaining radial coordinate is governed by a one-dimensional 
Schr\"odinger equation. Similar, the conformal expansion of distribution 
amplitudes in QCD aims to separate longitudinal degrees of freedom 
{}from  transverse ones. For the pion distribution amplitude it has 
a simple form\,\cite{exclusive}
\begin{eqnarray}
\phi_\pi(x,\mu) &=& 6x(1-x) \left\{1+
a_2(\mu_0)\left(\frac{\alpha_s(\mu)}{\alpha_s(\mu_0)}\right)^{50/81}
 C_2^{3/2}(2x-1)\right.
\nonumber\\&&\hspace*{2.3cm}{}\left.
+a_4(\mu_0) \left(\frac{\alpha_s(\mu)}{\alpha_s(\mu_0)}\right)^{364/405}
 C_4^{3/2}(2x-1)
+\ldots\right\}.
\end{eqnarray}  
All dependence on the longitudinal momentum fraction 
is included in Gegenbauer  polynomials $C_n^{3/2}$ which form 
an irreducible representation of the so-called collinear subgroup 
$SL(2,R)$ of the conformal group corresponding to M\"obius transformations 
on the light-cone, and the transverse-momentum dependence
(the scale-dependence) is governed by simple renormalization group equations:
The different partial waves,
labeled by different ``conformal spins'' $j=n+2$, do not mix with each other.
Since conformal invariance  is 
broken in QCD by quantum corrections, mixing of different  
conformal partial waves is absent to leading logarithmic accuracy only.
Still, conformal spin is a good quantum number in hard processes,
up to small corrections of order $\alpha_{s}^{2}$, and it is natural 
to expect that the hierarchy of contributions of different conformal 
partial waves is preserved at 
sufficiently low scales, meaning that only a few first ``harmonics'' 
are numerically important in B decays.  
This assumption is supported 
by the recent CLEO measurement of the $\pi\gamma^\ast\gamma $ form 
factor\,\cite{CLEO}   
which indicates that at scales of order 1 GeV the pion distribution amplitude
is already close to its asymtotic form $6 x(1-x)$.

Since the Gegenbauer polynomials  oscillate rapidly in high 
orders, their convolution with smooth functions
like in the r.h.s. of the sum rule (\ref{LCSR1}) is strongly suppressed.
{}For realistic values of the b quark mass it turns out that contributions 
of all polynomials with $n=4,6,\ldots$ are not important (unless the 
coefficients $a_n$ are abnormally large). The only significant potential 
correction to the ``S-wave'' contribution $6x(1-x)$ is with $n=2$. 
The parameter $a_2(1$~GeV) can be estimated from the CLEO data\,\cite{CLEO}
or from additional sum rules\,\cite{CZreport}. The (conservative) range
is $0<a_2(1$~GeV$)<0.5$. This uncertainty will eventually be eliminated 
when more high-precision data on exclusive processes involving pions become
available.   

\medskip
\noindent{\large\bf $B\to \pi e\bar\nu$: State of the Art}
\medskip 

The LCSR considered above has to be complemented by 
higher twist and radiative corrections.
Higher twist effects were calculated 
in Ref.\,\cite{Bel95} using the complete set of twist 3 and twist 4 
pion distribution amplitudes available from\,\cite{BF2}. The radiative 
correction was calculated very recently in Ref.\,\cite{KRWY,BBB}.   

The structure of the  radiative correction in the heavy quark limit 
is instructive and deserves to be mentioned here. The full 
expression is rather complicated, so I quote the answer\,\cite{BBB}
in the so-called local duality approximation corresponding to the 
limiting case
$M^2\to\infty$\footnote{The local duality limit has to be taken 
consistently with the heavy quark expansion, in particular the 
order of limits $m_b\to\infty$ and $M^2 \to \infty$ is important, 
see Ref.\,\cite{BBB} for the details.} in the sum rule:
\begin{eqnarray}
\lefteqn{
    \frac{f^{\rm stat}(m_b)}{f_\pi}[m_b^{3/2}f_+(0)]=}
\nonumber\\
    &=&-\omega_0^2\phi_\pi'(1,\mu)\Bigg[ 1+ \frac{\alpha_s}{\pi}C_f
    \Bigg(\frac{1+\pi^2}{4} +\ln\frac{m_b}{2\omega_0}  
   -\frac{1}{2}\ln^2\frac{m_b}{2\omega_0} 
    +\frac{1}{2}\ln \frac{2\omega_0}{\mu}
    \Bigg)\Bigg]  
\nonumber\\
&&{} -\omega_0^2 \frac{\alpha_s}{\pi}C_f
\left[ 
     \left(1-\ln\frac{2\omega_0}{\mu}\right)
   \int_0^1 dx \left(\frac{\phi_\pi(x)}{\bar x^2}+
\frac{\phi_\pi'(1)}{\bar x}\right)
     - \ln\frac{2\omega_0}{\mu}\int_0^1 dx \,\frac{\phi_\pi(x)}{\bar x}
\right]
\label{SR:hqlrad}
\end{eqnarray}  
where $\phi_\pi'(x) = (d/dx)\phi_\pi(x)$, $\bar x\equiv 1-x$
 and $\omega_0$ is the 
nonrelativistic continuum threshold $s_0 \simeq (m_b+\omega_0)^2$.
Local duality means that we identify the B meson with
a b quark accompanied by an arbitrary number of light quarks and gluons 
with total energy less than $\omega_0$ (in the b quark rest frame).

\begin{figure}
\centerline{\epsfxsize9.0cm\epsffile{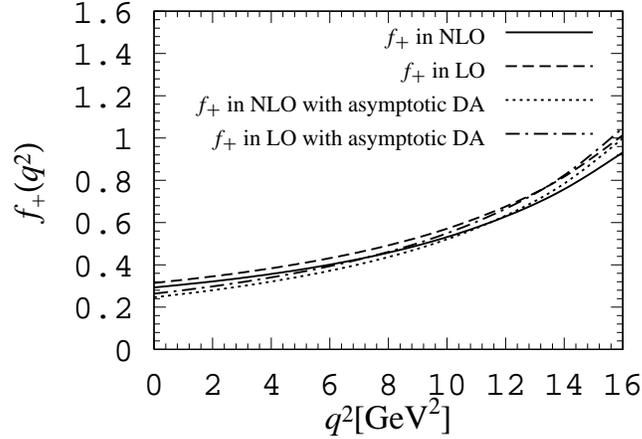}}
\caption[]{$f_+^{B\to\pi}(q^2)$ as a function of $q^2$ for two
  different sets of the leading twist pion distribution amplitude (DA). 
  The effect of including radiative corrections is a reduction of the
  form factor by about (4--7)\%.
}\label{fig:q2}
\end{figure}

Let us interpret the two pieces:
the first term on the right-hand side can be identified with the 
soft (end-point) contribution including the Born-term and its
radiative correction, while the second term corresponds to 
the usual mechanism\,\cite{hard} of hard gluon exchange.

The dependence on the collinear factorization scale $\mu$
must cancel the scale dependence of the pion distribution amplitude.
This implies that the structure of terms in $\ln \mu$ 
in the hard contribution is fixed by the structure of the leading order
soft term which is proportional to $\phi_\pi'(1,\mu)$.
Indeed, we find
\begin{eqnarray}
   \frac{d}{d \ln \mu} \phi_\pi'(1,\mu) &=& \frac{\alpha_s}{\pi}C_f
   \frac{d}{dx}\left[\int_0^1 dy\, V_0(x,y)\,\phi_\pi(y,\mu)\right]_{x\to 1} 
\nonumber\\
&=&{}-\frac{\alpha_s}{\pi}C_f\left\{
      \int_0^1 dx\,\left[\frac{\phi_\pi(x)+\bar x \phi_\pi'(1)}{\bar x^2} +
      \frac{\phi_\pi(x)}{\bar x}\right]-\frac{1}{2}\phi_\pi'(1)
                             \right\},
\label{scaledepend}
\end{eqnarray}
where $V_0(x,y)$ is the usual Brodsky-Lepage kernel, so that
the structure of $\ln \mu$ terms in (\ref{SR:hqlrad}) is reproduced. 
Note the subtraction term accompanying the naively divergent expression
$\int dx\, \phi_\pi(x,\mu)/\bar x^2$ \cite{hard}, which is similar to the
usual  ``plus'' prescription in the evolution kernel.
The lesson to be learnt is that LCSRs are fully consistent
with QCD and in fact can be used to study the factorization of hard
and soft (end-point) contributions.  

Some numerical results are shown in Fig.~2 and Fig.~3 \cite{BBB}.
It attracts attention that the radiative correction is small, at most 7\%
in the whole $q^2$ range, and the higher twist effects 
appear to be under control\footnote{
The large twist 3 correction is exactly calculable in terms of the 
quark condensate.}. Possible deviation of the pion distribution amplitude 
from its asymptotic form mainly affects the slope of the form factor and
has little impact on the normalization. The corresponding spectrum 
$dB/d q^2$ has to increase somewhat from $q^2=0$ to  
$q^2 \leq 15$ GeV$^2$ if the pion distribution amplitude is close 
to its asymptotic expression, and it decreases with $q^2$ if the distribution
amplitude has large corrections\,\cite{CZreport}, see Fig.~3 in 
Ref.\,\cite{BBB}.
This behavior can be checked experimentally in the near future.   

\begin{figure}
\centerline{\epsfxsize9.0cm\epsffile{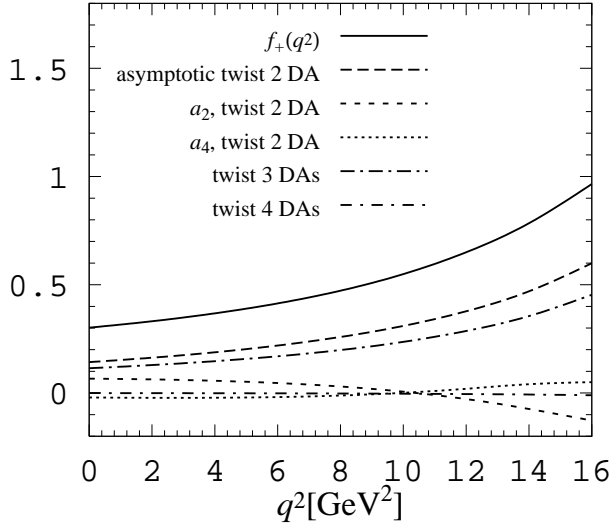}}
\caption[]{The several contributions to the light-cone sum rule for the form
  factor $f_+^{B\to\pi}(q^2)$
  as a function of $q^2$, using the leading twist distribution
  amplitude of {\protect\cite{BF1}}.}\label{fig:contr}
\end{figure}

The analysis of theoretical uncertainties in the sum rule method is a
difficult issue in general. Using state-of-the-art LCSRs
 including radiative corrections and 
higher twist effects up to twist 4, and with some
better knowledge of $m_b$ and $f_B$, one can expect a theoretical accuracy
up to 10\% in form factors which translates to 20\% uncertainty in the 
decay rates. Yet higher accuracy is not feasible within the sum rule 
method. 

\medskip
\noindent{\large\bf Other Heavy-to-Light Decays}
\medskip

Apart from the simplest process  $B\to\pi e\bar\nu$ which historically 
attracted most of the attention, LCSRs have been derived 
for semileptonic $B\to\rho e\bar\nu$ decays\,\cite{ABS,BB97}, see 
{}Fig.~4, and for 
rare radiative decays induced by flavor-changing neutral currents,
most notably $B\to K^*\gamma$ \cite{ABS}.
 Other decays studied are $B_s\to K^*\gamma$, 
$B_u\to\rho(\omega)\gamma $ and $B_s\to\phi\gamma$ \cite{ABS}.
In addition, the $B\to K^* l^+l^-$, $B\to K l^+l^-$
decay form factors  have been calculated using
the light-cone approach in Ref.\,\cite{ALIEV96,ALIEV97}.
The relevant form factors are too numerous to be presented here.

\begin{figure}
\centerline{\epsfxsize8.5cm\epsffile{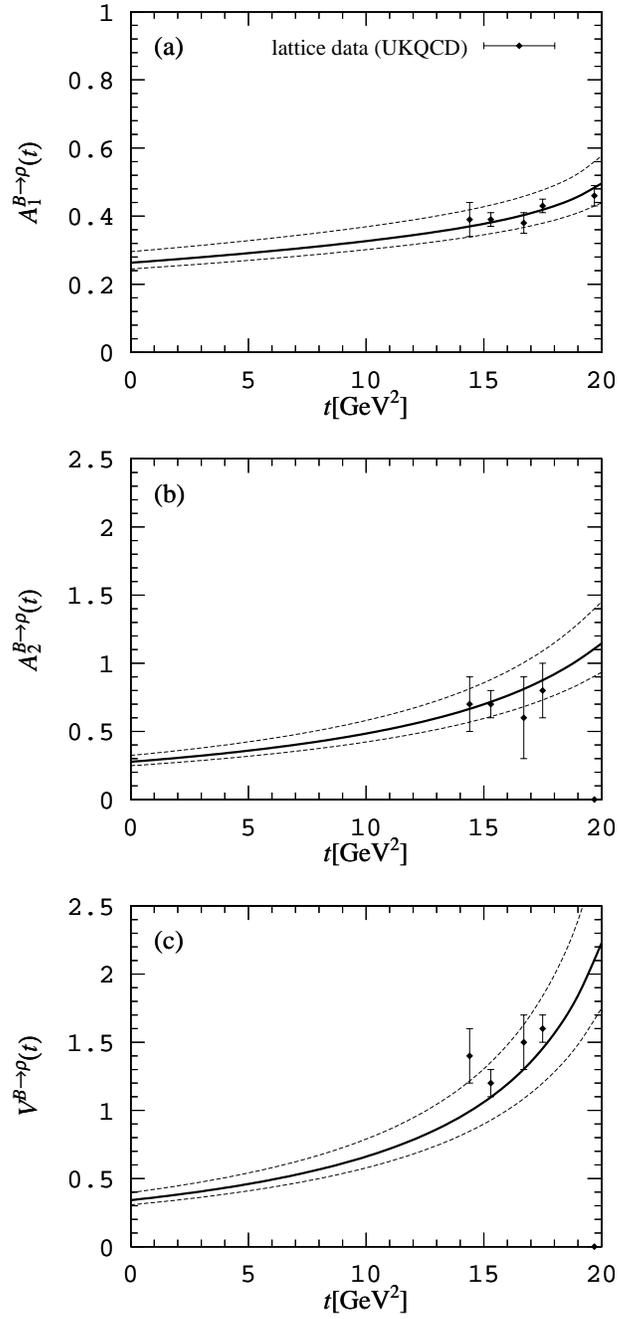}}
\caption[]{Semileptonic $B\to\rho e\bar\nu$ decay form factors 
as functions of $t=q^2$  
from Ref.\,\protect{\cite{BB97}}. The dashed
lines give error estimates. The data points are from 
Ref.\,\protect{\cite{Flynn96}}.}
\end{figure}

The results of\,\cite{ABS,ALIEV96} have eventually to be updated
to include radiative corrections and using
the revised distribution amplitudes of the vector 
meson\,\cite{BB96} and surface terms for  
the continuum subtraction\,\cite{BB97}. 
In the analysis of \,\cite{ALIEV96} one should also take into 
account $SU(3)$ violating asymmetry in the K meson  
distribution amplitude.
I do not expect significant numerical changes, however.  

The LCSR method can be  used to estimate the 
long-distance contributions of four-fermion operators 
to the decay $B^+\to\rho^+\gamma$ \cite{KHO95,ALI95} which appear to be 
of order 20\% of the short-distance contribution to the decay rate. 
The same approach was applied to the decay $B\to \mu \bar\nu_\mu\gamma$ 
in Ref.\,\cite{KHO95,EIL95}.

\medskip
\noindent{\large\bf Other Applications}
\medskip

My discussion so far was concentrated on the B decays which are topical 
for this conference. The LCSR approach is, however, 
quite general and is equally useful for form factors of light hadrons,
where it has the similar advantages of being applicable in a wide range 
of momentum transfers and using simpler dispersion relations.
Sample applications include the (electromagnetic) pion form factor
\cite{BH94}, $\pi A_1\gamma$ form factor\,\cite{Bel95a}, $\gamma^*\rho\to\pi$
\cite{BH94,Kho97} and $\gamma^*\gamma\to \pi^0$ \cite{Kho97} transitions.

\medskip
\noindent{\large\bf Light-Cone Sum Rules for Hadron Matrix Elements}
\medskip

A conceptually similar but technically somewhat more complicated 
modification of the LCSR approach is useful for 
calculations of matrix elements of local operators between hadron 
states (like baryon magnetic moments) or  involving two heavy and one 
light hadron, like $D^*\to D\pi$ decays. In this type of problems there is no
large scale involved (except, possibly, a heavy quark mass) and the 
light-cone approach has to be compared with the method  of
Ref.\,\cite{bilocal}
with explicit separation of local and bilocal power corrections. 
The work Ref.\,\cite{Bel95} contains a
rather detailed introduction to the LCSR technique in this context,
which is more readable than the original papers\,\cite{BBK,BF1}.  

To give an example, I will consider calculation of the $g_{BB^*\pi}$
coupling. The starting point in both approaches is the same correlation
function (\ref{B-Pi}) where the vector current now serves as an interpolating
field for the $B^*$ meson and it is therefore convenient to change 
the notation for the corresponding momentum $q\to p_{B^*}$:
\begin{equation}
\Pi_\mu(p_B^2,p_{B^*}^2) = i\int\!d^4z\,e^{-ip_Bz}
\langle \pi(p_\pi)|T\{\bar u(0)\gamma_\mu b(0) b(z)i\gamma_5 d(z)\}|0\rangle.
\label{CFgBBpi}
\end{equation}
As explained in detail in Ref.\,\cite{Bel95}, in order to apply the 
short-distance expansion to this correlation function one has to 
take the soft pion limit $p_\pi\to 0$ so that 
$p_B=p_{B^*}=p$. Therefore, a double Borel transformation cannot be applied 
and nondiagonal transitions from ground to excited 
states produce a single-pole contribution to (\ref{CFgBBpi}): 
$$ \frac{g_{BB^*\pi}}{(p^2-m_B^2)^2} + A\cdot \frac{1}{(p^2-m_B^2)}.$$
The constant $A$ creeps into the sum rule, which has the following 
schematic structure\,\cite{EK85}:
\begin{equation}
   \#\cdot g_{BB^*\pi} + M^2 A = m_b f_\pi M^2 \exp\left[
      \frac{m_{B^*}^2-m_b^2}{2M^2}+\frac{m_{B}^2-m_b^2}{2M^2}\right]
      +\ldots
\end{equation}
Thus, one sum rule has to be used to determine two unknown constants ---
$g_{BB^*\pi}$ and $A$ --- which reduces the accuracy. In addition,
it is in principle not possible to keep the B and B$^*$ masses different
from each other, since, as I already mentioned in the introduction, 
in this case the double-pole term is not present in the correlation 
function. This deficiency is marginal for the case in question, 
but it can be crucial in other applications.
Historically, the need to take into account the mass
difference of the proton and $\Sigma$-hyperon in the weak decay 
$\Sigma\to p\gamma$ \cite{BBK} has been  the prime motivation for 
the development of the LCSR approach.

As I emphasized already, the  main characteristic feature of 
LCSRs is that short-distance expansion is replaced 
by expansion in powers of the deviation from the light-cone (or transverse 
distance in light-cone coordinates). 
The light-cone  expansion corresponds to a more general kinematics, with 
the pion being on-shell $p_\pi^2=m_\pi^2\simeq 0$
but with nonzero momentum, so that $2(p_\pi\cdot p_B) = p_{B}^2-p_{B^*}^2$ 
can be arbitrary large and one can take $p_B^2$ and $p_{B^*}^2$ as two
independent variables. Taking the Borel transform in both of them
one obtains the sum rule\,\cite{Bel95} 
\begin{equation}
   \#\cdot g_{BB^*\pi}  = m_b f_\pi \phi_\pi(x) \exp\left[
      \frac{m_{B^*}^2-m_b^2}{2M^2_1}+\frac{m_{B}^2-m_b^2}{2M^2_2}\right]
      +\ldots
\end{equation}
where the argument of the pion distribution amplitude is fixed
by the ratio of the two Borel parameters
\begin{equation}
   x = \frac{M_1^2}{M_1^2+M_2^2}\sim \frac{m_{B^*}^2}{m_{B^*}^2+m_{B}^2}
   \simeq \frac{1}{2}. 
\end{equation}   
The premium is that single-pole terms are absent and one can keep 
$m_{B^*}\neq m_B$, while the price to pay is that one has a nontrivial 
new input:
The pion distribution amplitude in approximately the middle point.
Lacking direct experimental measurement of $\phi_\pi(1/2)$ one 
can consider this quantity as a nonperturbative parameter to be 
found from one suitable sum rule and used elsewhere\,\cite{BF1}, 
similar to the usual way how the gluon condensate is determined and
used in the sum rules. 
Note that this quantity is not related to a matrix element of any 
local operator, which illustrates that the sum rule is not related 
to a short distance expansion. The dedicated study in Ref.\,\cite{BF1} 
resulted in the estimate $\phi_\pi(1/2) = 1.2\pm 0.2$ which is only slightly
below the asymptotic value $\phi_\pi(1/2) = 3/2$.

The same approach is applicable to the calculation of amplitudes involving
emission of a real photon with the advantage that  
the  photon distribution amplitudes
are expected to be very close to their asymptotic form\,\cite{BBK}.
As the result, for photon radiation there are no free parameters
and it was checked that the LCSR approach works very well for the proton
and neutron magnetic moments\,\cite{BF1}. Other applications 
include: $\Sigma\to p\gamma$ decay\,\cite{BBK}, 
$g_{\pi NN}$ and $g_{\rho\omega\pi}$
couplings\,\cite{BF1} and the radiative decays 
$D^*\to D\gamma$, $B^*\to B\gamma$ \cite{BBgamma}.

The LCSR result is $g_{B^*B\pi}= 29\pm 3 $~\cite{Bel95} 
with an error corresponding to the estimated theoretical uncertainty. 
In the same framework, the strong coupling constants of the scalar 
and axial B mesons with the pion have been estimated yielding
the following predictions for the observable 
strong decay widths\,\cite{COL95}:
$\Gamma( B(0^{++}) \to B \pi)\simeq \Gamma( B(1^{++}) 
\to B^* \pi)\simeq 360 ~\mbox{MeV}$. An analogous method was used
in Ref.\,\cite{AlievBrho} to obtain the $BB^*\rho$ coupling. 
 
\medskip
\noindent{\large\bf Summary and Further Prospects}
\medskip

I have given a short introduction to the technique of light-cone sum rules,
their theoretical  background and main modifications. This approach is a
derivative of the QCD sum rule method\,\cite{SVZ} and combines
characteristic features of sum rules with the theory of hard exclusive 
processes. Main idea and the defining feature of a generic 
LCSR is that the short-distance Wilson operator product 
expansion is substituted
by the light-cone expansion in operators of increasing twist;  
for given twist the expansion in local operators is replaced by the expansion 
in conformal partial waves. Each term in the partial wave 
 expansion is well defined and has the expected 
asymptotic behavior at the end-points of the 
phase-space. The approach involves an implicit  physical assumption that the 
conformal spin presents a ``good'' approximate quantum number in hard
exclusive processes in QCD, and 
it is this physics issue that will eventually be decided by 
the success (or failure) of the LCSR programm.

Although the approach is already 10 years old, full understanding of 
its advantages and potential is rather recent.
LCSRs  
can be used for a broad range of processes from which I mainly discussed 
applications to heavy quark decays. There is  room for further
improvement: The existing LCSRs are in most cases
derived to leading twist accuracy only and do not include radiative 
corrections. A few methodical questions need to be clarified as well.

Main input in the sum rules is provided by hadron light-cone distribution
amplitudes. They have a direct physical interpretation and in this 
sense are as basic as conventional parton distributions. 
{}For a further progress in LCSR calculations  a systematic 
study of distribution amplitudes is mandatory. The present situation is 
not satisfactory and requires both theoretical and experimental efforts.
The leading twist distributions can be studied experimentally and 
there is increasing evidence that they are not far from their asymptotic 
{}form (the 'S-wave' contribution to the conformal partial wave expansion).
Several works on higher-twist meson distribution amplitudes are in progress.
Results on higher-twist baryon distributions are so far absent and 
would be most welcome.

\medskip
\noindent{\large\bf Acknowledgements}
\medskip

It is a pleasure for me to thank A.~Ali, I.~Balitsky, P.~Ball, 
V.M.~Belyaev, H.G.~Dosch, I.~Filyanov (Halperin), 
A.~Khodjamirian, R.~R\"uckl and H.~Simma for a very rewarding collaboration
on subjects related to this report. Special thanks are due to the 
organizers of this conference for the invitation and hospitality.

\end{document}